\documentclass[a4paper,fleqn,usenatbib]{mnras}

\usepackage{graphicx}
\usepackage{natbib}
\usepackage{hyperref}
\usepackage{amsmath}
\usepackage{amssymb}

\defcitealias{Prole2021}{P21}
\defcitealias{Danieli2019}{Danieli \& van Dokkum 2019} 
\defcitealias{Carleton2019}{Carleton et al. 2019} 

\newcommand{\RNum}[1]{\uppercase\expandafter{\romannumeral #1\relax}}

\newcommand{\emcee}{{\scshape emcee}}

\newcommand{\recObs}{\ensuremath{\bar{r}_{e}}}
\newcommand{\uaeObs}{\ensuremath{\bar{\mu}_{e}}}
\newcommand{\colourObs}{\ensuremath{\bar{c}}}

\newcommand{\recObsIndexed}{\ensuremath{\bar{r}_{e, i}}}
\newcommand{\uaeObsIndexed}{\ensuremath{\bar{\mu}_{e, i}}}
\newcommand{\colourObsIndexed}{\ensuremath{\bar{c}_{i}}}

\newcommand{\recPhys}{\ensuremath{\hat{r}_{e}}}
\newcommand{\colourRest}{\ensuremath{\hat{c}}}
\newcommand{\recObsInt}{\ensuremath{\tilde{r}_{e}}}
\newcommand{\uaeObsInt}{\ensuremath{\tilde{\mu}_{e}}}

\newcommand{\Mstar}{\ensuremath{M_{*}}}
\newcommand{\redshift}{\ensuremath{z}}

\newcommand{\Msun}{\ensuremath{\mathrm{M_{\odot}}}}

\newcommand{\selectionRec}{3\arcsec$\leq$\recObs$<$10\arcsec}
\newcommand{\selectionUae}{24$\leq$\uaeObs$<$27}
\newcommand{\selectionColour}{0$\leq$\colourObs$<$0.42}

\newcommand{\githuburl}{\texttt{https://github.com/danjampro/udg-sizes}}
\newcommand{\numObserved}{188}

\newcommand{\emceeBurnin}{1000}
\newcommand{\emceeNsamples}{2 million}
\newcommand{\gridRangeAlpha}{-1.65$\leq$$\alpha$$<$1.25}
\newcommand{\gridStepAlpha}{0.025}
\newcommand{\gridRangeBeta}{0.2$\leq$$\beta$$<$0.8}
\newcommand{\gridStepBeta}{0.025}

\newcommand{\modelHyperPars}{\ensuremath{\boldsymbol{\theta}}}
\newcommand{\modelPhysPars}{\Mstar, \redshift, \recPhys, \colourRest}
\newcommand{\modelObsPars}{\uaeObs, \recObs, \colourObs}
\newcommand{\modelObsParsIndexed}{\uaeObsIndexed, \recObsIndexed, \colourObsIndexed}

\newcommand{\resultBeta}{$\beta$=0.40$\pm$0.04}
\newcommand{\resultBetaFinal}{$\beta$=0.40$\pm$0.07}
\newcommand{\resultAlpha}{$\alpha$=-1.44$\pm$0.05}
\newcommand{\resultMstarRange}{$10^{7}$$\leq$\Mstar/\Msun$\leq$$10^{9}$}
\newcommand{\resultUdgPower}{\ensuremath{3.54\pm0.33}}
\newcommand{\resultDwarfPercentage}{\ensuremath{78\pm5\%}}
\newcommand{\resultUDGPercentage}{\ensuremath{84\pm1\%}}

\author[D. J. Prole et al.]{D. J. Prole$^{1, 2}$\thanks{daniel.prole@mq.edu.au} \\
	$^{1}$Research Centre for Astronomy, Astrophysics \& Astrophotonics, Macquarie University,
	Sydney, NSW 2109, Australia \\
	$^{2}$Department of Physics \& Astronomy, Macquarie University, Sydney, NSW 2109, Australia \\
}

\title[The size distribution of LSB galaxies]{The stellar mass - physical effective radius relation for dwarf galaxies in low-density environments}

\date{}

\pubyear{2018}

\begin{document}
	\label{firstpage}
	\pagerange{\pageref{firstpage}--\pageref{lastpage}}
	\maketitle

\begin{abstract}
	
The scaling relation between stellar mass (\Mstar) and physical effective radius ($r_{e}$) has been well-studied using wide spectroscopic surveys. However, these surveys suffer from severe surface brightness incompleteness in the dwarf galaxy regime, where the relation is poorly constrained. In this study, I use a Bayesian empirical model to constrain the power-law exponent $\beta$ of the \Mstar-$r_{e}$ relation for late-type dwarfs (\resultMstarRange) using a sample of \numObserved\ isolated low surface brightness (LSB) galaxies, accounting for observational incompleteness. Surprisingly, the best-fitting model (\resultBetaFinal) indicates that the relation is significantly steeper than would be expected from extrapolating canonical models into the dwarf galaxy regime. Nevertheless, the best fitting \Mstar-$r_{e}$ relation closely follows the distribution of known dwarf galaxies. These results indicate that extrapolated canonical models over-predict the number of large dwarf (i.e. LSB) galaxies, including ultra-diffuse galaxies (UDGs), explaining why they are over-produced by some semi-analytic models. The best-fitting model also constrains the power-law exponent of the physical size distribution of UDGs to $n\mathrm{[dex^{-1}]}\propto$$~r_{e}^{\resultUdgPower}$, consistent to within 1$\sigma$ of the corresponding value in cluster environments and with the theoretical scenario in which UDGs occupy the high-spin tail of the normal dwarf galaxy population.

\end{abstract}


\section{Introduction}

The advent of wide-field spectroscopic surveys has allowed systematic distance measurements of millions for galaxies, thus enabling the scaling relation between stellar mass (\Mstar) and physical effective radius to be determined empirically \citep{Shen2003, vanderWel2014, Lange2015}. This relation plays an important role in understanding galaxy formation theoretically, forming a basic ingredient of semi-analytical models \citep[e.g.][]{Somerville1999} as well as a basis of comparison for cosmological simulations \citep[e.g.][]{Buck2020, Davison2020}. However, spectroscopic surveys invariably suffer from completeness effects, meaning the relation is poorly constrained in the dwarf galaxy regime (\Mstar$\leq$$10^{9}$\Msun) where galaxies typically have lower surface brightness \citep{Martin2019}. 

\indent Low surface brightness galaxies (LSBGs), characterised by surface brightness levels fainter than 24 magnitudes per square-arcsecond averaged within their effective radii, are often completely missed in spectroscopic surveys \citep{Wright2017}. These galaxies are uncommonly large for their stellar masses, the most extreme of which ($r_{e}$$\gtrsim$1.5 kpc) being known as ``ultra-diffuse galaxies" \citep[UDGs, ][]{vanDokkum2015}.

LSBGs can form and evolve through a variety of processes \citep[e.g.][]{Jackson2020}, which can be either secular (proceeding without external influence) or environmentally-driven. Secular channels are more influential for field galaxy populations owing to the decreased environmental density. One example is supernovae feedback, which can enlarge normal dwarf galaxies, thereby turning them into LSBGs \citep{DiCintio2017}. Higher than average halo spin parameters can also create dwarf galaxies with large sizes \citep{Amorisco2016, Rong2017, Tremmel2020}. One prediction from this scenario is that the distribution of effective radii among the field LSBG population should follow a power-law with an exponent similar to that measured in cluster environments \citep{vanderBurg2016}. Recent kinematic studies have indicated that the theory is viable for UDGs in lower density environments \citep{Pina2020}. However, a systematic study of this field population has yet to be performed.

Distances to LSBGs in nearby galaxy groups and clusters can be estimated by associating them with their environment. Prohibitive difficulties in measuring distances to more isolated galaxies \citep[e.g.][]{Trujillo2019} have caused the field population to remain poorly understood. This problem will not be fully addressed by the next generation of all-sky extragalactic spectroscopic surveys (e.g. from 4MOST), which will not provide spectroscopic distance estimates for large samples of LSB galaxies because they are  too faint to meet survey selection criteria. Other techniques such as surface brightness fluctuations \citep{Greco2020} will not be useful for galaxies far outside of the local group.

In this paper, I use an empirical model to constrain the stellar mass - effective radius relation in the dwarf galaxy regime using a method that implicitly accounts for surface brightness incompleteness. This is achieved by fitting the model to the sample of field LSBGs recently obtained by \cite{Prole2021} (hereafter P21).  I investigate whether the model is consistent with the theoretical predictions for UDGs from \cite{Amorisco2016} and corresponding measurements inside galaxy clusters \citep{vanderBurg2016} as well as groups \citep{vanderBurg2017}. While no distance measurements are available for the observed sample, the model overcomes this problem by assuming the galaxies are distributed smoothly along with the matter in the Universe.

Throughout this paper, surface brightnesses are always given in units of magnitudes per square arc-second and the AB magnitude system is used. $\Lambda$CDM cosmology is assumed, with $\Omega_{\mathrm{m}}$=0.3, $\Omega_{\Lambda}$=0.7 and H$_{0}$=70 kms$^{-1}$Mpc$^{-1}$.

\vspace*{-5mm}
\section{Data}
\label{section:data}

The publicly-available LSBG catalogue from \citetalias{Prole2021} is used in this study. In particular, the subset of blue\footnote{Instead of selecting blue LSBGs in the colour-S\'ersic index plane as was done in \citetalias{Prole2021}, I make the simplified selection of $(g-r)$$<$0.42. This simplifies the analysis because the S\'ersic index can be neglected from the model.} LSBGs are selected because they show no correlation with local structure and may be considered to exist in low-density environments. These galaxies were identified in deep optical imaging from HSC-SSP PDR2 \citep{Aihara2019} and have $r$-band structural parameter measurements based on single component 2D S\'ersic fits. The sample are selected to have \selectionUae\ and \selectionRec, where \uaeObs\ and \recObs\ are the \textit{measured} apparent surface brightness averaged within the effective radius and the angular circularised effective radius respectively. A further selection cut is applied in colour: \selectionColour, where \colourObs\ is the observer-frame $(g-r)$ colour in magnitudes. These criteria leave \numObserved\ galaxies in the sample. The maximum S\'ersic index ($n$) of the sample is 2.2. The sample are likely to be local ($z$$<$0.2), with the main bulk of sources being much closer ($z$$<$0.1). For a full description of the sample, selection criteria and distance arguments, the reader is referred to \citetalias{Prole2021}. 

\vspace*{-5mm}
\section{Model}
\label{section:model}


The objective of the statistical model is to reproduce the joint distribution of observed quantities from the \citetalias{Prole2021} sample given a set of model hyper parameters \modelHyperPars:
\begin{equation}
p_{obs}(\modelObsPars~|\modelHyperPars)
\end{equation}

In order to achieve this, we must model the underlying joint distribution of physical properties that are $observed$ given the selection criteria of \citetalias{Prole2021}:
\begin{equation}
p_{obs}(\modelPhysPars~| \modelHyperPars) \propto \epsilon(\modelPhysPars) p(\modelPhysPars~| \boldsymbol{\theta})
\label{equation:model}
\end{equation}

where  \Mstar\ is the stellar mass, \redshift\ is the redshift, \recPhys\ is the physical circularised effective radius and \colourRest\ is the rest-frame $(g-r)$ colour. $\epsilon$ is the fraction of recovered galaxies given the selection criteria used in \citetalias{Prole2021}, parametrised as a function of \uaeObs, \recObs.

\indent Empirical relations pertaining to late-type galaxies are used throughout the model. This is justified because $\epsilon$=0 for galaxies that do not satisfy the selection criteria, and the selection criteria of \citetalias{Prole2021} target the late type population. The reader is referred to the selection criteria of ``late-type" galaxies discussed in \cite{Baldry2012} and \cite{Lange2015} which are consistent with the sample in this study (blue colours and $n$$<$2.5).

\subsection{Redshift distribution}

The results of \citetalias{Prole2021} showed that the observed LSBG sample do not correlate spatially with local structure.
I therefore assume that the spatial distribution of these galaxies can be expected, when averaged over large-enough areas, to follow the mean cosmological distribution of mass within the volume element, such that:
\begin{equation}
p(z) \propto \Omega_{m}(z)\rho_{\mathrm{crit}}(z)\frac{dV}{dz}
\label{equation:redshift}
\end{equation}

where $\Omega_{m}$ is the matter-density parameter, $\rho_{\mathrm{crit}}$ is the critical density of the Universe and $\frac{dV}{dz}$ is a term that is proportional to the cosmological volume element.

\indent I also make assumption that the redshift distribution $p(z)$ is independently distributed from the physical properties of the galaxies. This is justified if the observed sample are isolated and local enough such that any redshift evolution is negligible. Therefore, it is possible to write:
\begin{equation}
p(\modelPhysPars) = p(\Mstar, \recPhys, \colourRest)p(z)
\end{equation}

\subsection{Stellar mass distribution}

The galaxy stellar mass function (SMF) is well-constrained for late-type galaxies well into the dwarf galaxy regime \citep{Baldry2012, Sedgwick2019}. The SMF is commonly parametrised using a Schechter function:
\begin{equation}
p(\Mstar|M_{0}, \alpha) \propto  \mathrm{exp} \left(\frac{-M_{*}}{M_{0}}\right) \left(\frac{M_{*}}{M_{0}}\right)^{\alpha}
\end{equation}

where $M_{0}$ is the ``mass-break" parameter and $\alpha$ is the ``faint end slope" which governs the relative abundance of low-mass galaxies. Fiducial values are $ \alpha=-1.45$, $M_{0}=10.56\Msun$ \citep{Sedgwick2019b}. In this work, $M_{0}$ is fixed to this value, resulting in a single free parameter, $\alpha$, to govern the SMF. 

\subsection{Stellar mass - effective radius relation}

\indent The scaling relation between stellar mass and physical size for late-type galaxies has been well-constrained for stellar masses greater than around $10^{9}$\Msun\ \citep[][figure \ref{figure:mstar-rec}]{Shen2003, vanderWel2014, Lange2015}. I adopt the results from \cite{Shen2003} as they fit \textit{circularised} effective radii as opposed to semi-major effective-radii, while also fitting for the intrinsic scatter in the relation.

\indent An interesting property of this model is that it reduces to a simple form for galaxies with stellar masses lower than $\sim$$10^{10}\Msun$: \recPhys$\propto$\Mstar$^{\gamma}$, $\sigma_{\ln{\recPhys}}$=$const$, where $\sigma_{\ln{\recPhys}}$ quantifies the log-normal scatter in the relation. However, this model is not constrained for galaxies less massive than \Mstar$\sim$$10^{9}$\Msun\ owing to incompleteness in the observed sample used to measure it. I therefore consider a simple power-law extension to the \cite{Shen2003} model for galaxies lower than this limit:
\begin{equation}
\log(\recPhys, \Mstar<10^{9}\Msun) = a (\Mstar / \Msun)^{\beta}
\end{equation}



where $\beta$ is the power law exponent that governs the sizes of dwarf galaxies at a given stellar mass, and $a$ is chosen to keep the \Mstar-\recPhys\ relation continuous\footnote{It is necessary for the relation to be kept continuous because surface brightness incompleteness is negligible at stellar masses higher than $\sim$1$0^{9}$\Msun \citep[e.g.][]{Wright2017}.}. $\beta$  is left as a free parameter of the model. Setting $\beta$=$\gamma$=0.14 is equivalent to an extrapolation of the \cite{Shen2003} relation. I make the assumption that $\sigma_{\ln{\recPhys}}$=$const$ can be extrapolated to lower masses than $10^{9}$\Msun.

\subsection{Rest frame colours}
\label{section:modelColours}

The rest-frame colours of galaxies correlate with their stellar masses; higher mass galaxies are typically redder. This is an important effect to model because of the colour selection used in \citetalias{Prole2021}. I adopt the empirical relation between rest-frame ($g-r$) colour and stellar mass using data obtained for the GAMA survey \citep{Taylor2011}. The data are well-modelled by a mean trend plus constant scatter term ($\sim$$0.1$ mag). However, for low stellar masses ($<10^{9}$\Msun), the GAMA sample suffers from surface brightness incompleteness \citep{Wright2017}. This is important because higher surface brightness dwarf galaxies are typically bluer than their LSB counterparts. I make the assumption that the rest-frame ($g-r$) colours of dwarf galaxies can be modelled by a normal distribution with mean 0.35 and a standard deviation of $\sim$$0.1$ mag, in agreement with the sample of \citetalias{Prole2021} as well as the field UDGs of \cite{Leisman2017}. The stellar mass-colour relation is naturally continuous given these assumptions. The importance of this assumption is discussed in $\S$\ref{section:results}.


\subsection{Sampling}

It is not easy to evaluate the model  (equation \ref{equation:model}) directly because of its high-dimensionality and non-parametric description of measurement errors (see $\S$\ref{section:error}). The alternative approach adopted here is to sample from the model using Marcov chain Monte Carlo (MCMC). Specifically, ensemble sampling via \emcee\ is used \citep{Foreman-Mackey2013}. Given a set of model hyper parameters \modelHyperPars=($\alpha$, $\beta$), 200 walkers are used to obtain \emceeNsamples\ total samples of (\modelPhysPars), after discarding the initial \emceeBurnin\ samples from each walker. The model is adequately sampled given these settings: the sampling chain runs for $\sim$200 times the autocorrelation time for each parameter. No hard parameter limits are imposed during sampling (other than the selection cuts discussed in $\S$\ref{section:data}) since the recovery fraction $\epsilon$ naturally imposes such limits. 

\subsection{Distance projection}
\label{section:projection}

The model needs to convert physical quantities (\modelPhysPars) into observables (\modelObsPars) in order to compare with observations and to evaluate the recovery fraction $\epsilon$. Physical sizes are projected to angular sizes using the cosmological angular diameter distance. Stellar masses are converted into absolute magnitudes as a function of rest-frame colour and stellar mass using the results of  \cite{Taylor2011}. Cosmological surface brightness dimming is accounted for by using the luminosity distance in combination wit the angular diameter distance when projecting in redshift. $k$-corrections are assigned to model samples based on their redshift and rest-frame colour in accordance with \cite{Chilingarian2010}.

\subsection{Measurement error}
\label{section:error}

Measurement errors are modelled using the artificial source injections from \citetalias{Prole2021}: The catalogue is binned two-dimensionally in \uaeObsInt, \recObsInt, the intrinsic (free from measurement error) observer-frame surface brightness and angular, circularised effective radii. Model samples are assigned relative errors in these quantities by directly assigning them a fractional error from a random artificial source in the same bin. This approach is non-parametric and accounts for both random and systematic uncertainties as a function of \uaeObsInt, \recObsInt. The ability to include the effects of measurement error this way is one of the key advantages of sampling the model rather than evaluating it analytically.

\section{Results}
\label{section:results}

All three observed quantities (\uaeObs, \recObs and \colourObs) are used to fit the model to the data. This prompts a 3D likelihood term of the form:
\begin{equation}
\mathcal{L}(\modelHyperPars) = \prod_{i} p(\modelObsParsIndexed~|\modelHyperPars)
\label{equation:likelihood}
\end{equation}

\noindent where $i$ loops over the observations. The probability density $p(\modelObsParsIndexed~|\modelHyperPars)$ is estimated from the model samples using a 3D Gaussian kernel. Since kernel density estimation (KDE) performs poorly on bounded problems (i.e. selection cuts) with heavily skewed distributions, as is the case here, a box-cox transform \citep{Box1964} is first applied to both the observations and model samples. This improves the ability of the KDE to represent the probability density at the selection boundaries.

Sampling \emceeNsamples\ times from equation \ref{equation:model} is computationally demanding. This means that it is not practical to use a MCMC approach to evaluate equation \ref{equation:likelihood} for many sets of \modelHyperPars. I therefore adopt a grid-based approach for exploring the likelihood space. In this method, a uniformly sampled grid of hyper parameters \modelHyperPars$_{i, j}$ is considered, with the following range: \gridRangeAlpha\ in steps of \gridStepAlpha, and \gridRangeBeta\ in steps of \gridStepBeta.

\begin{figure}
	\includegraphics[width=0.8\linewidth]{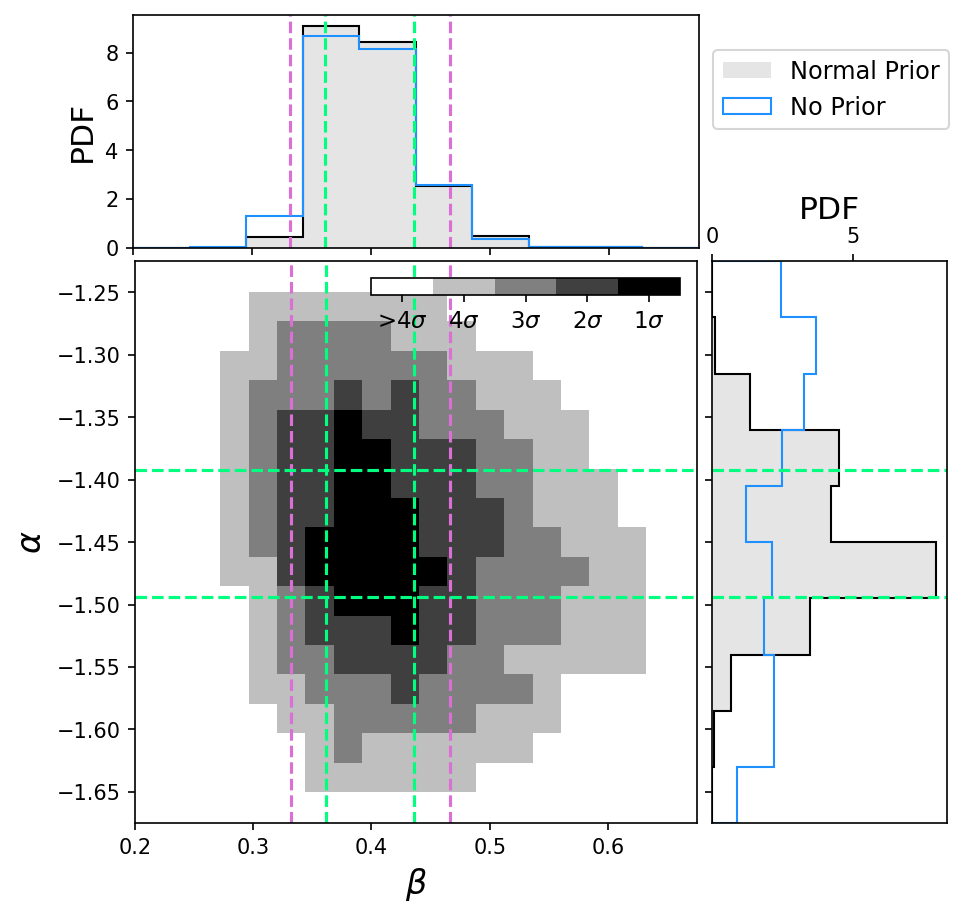}
	\centering
	\vspace*{-3mm}
	\caption{The posterior distribution as a function of hyper parameters $\alpha$ (faint end slope of the SMF) and $\beta$ (power-law exponent of the \Mstar-\recPhys\ relation in the dwarf galaxy regime). The marginal posteriors are shown for a uniform (blue histograms) and normal prior (grey histograms). The marginalised $\pm$1$\sigma$ intervals are shown in green. The pink lines show the uncertainty on $\beta$ after taking additional systematic uncertainties into account.}
	\label{figure:corner}
\end{figure}

The faint end slope the SMF as measured by \cite{Sedgwick2019b} is well-constrained and includes the contribution from the LSBG population over the mass range considered here. This motivates a normal prior on $\alpha$ with mean a mean of -1.45 and standard deviation of 0.05 that is multiplied with the likelihood to form the posterior distribution. The prior is found to be influential for constraining $\alpha$, which is otherwise unconstrained by the observations (figure \ref{figure:corner}). However, it does not significantly impact the marginalised posterior for $\beta$ as compared to a uniform prior over the fitting range.

\begin{figure}
	\includegraphics[width=0.85\linewidth]{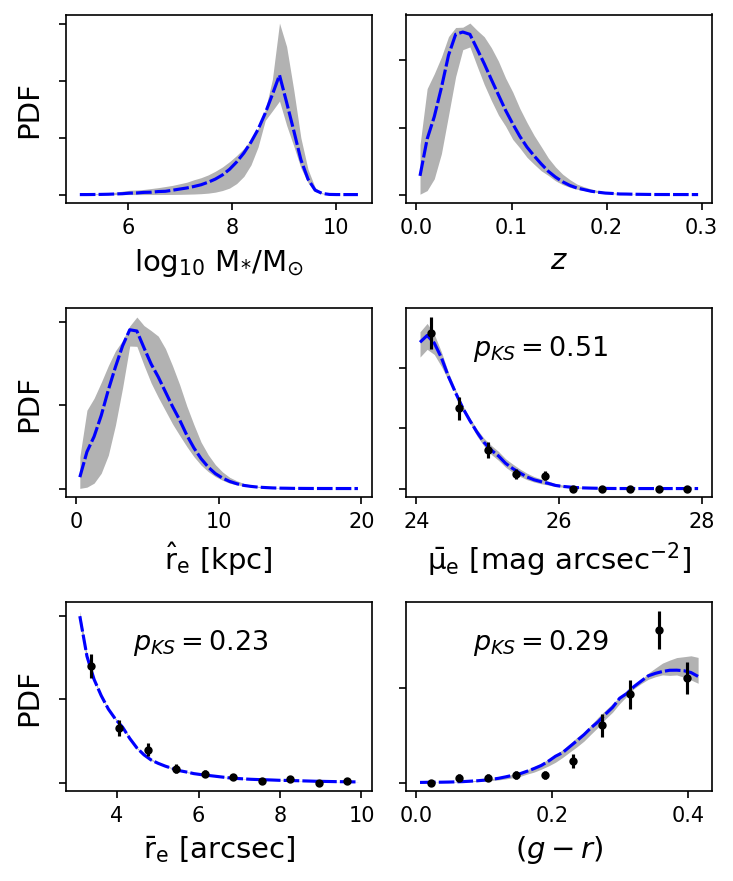}
	\centering
	\vspace*{-2mm}
	\caption{Predicted observations from the best-fitting model (blue lines) and 90\% confidence interval (grey regions) vs the observations (black points) given the selection efficiency from the \protect\citetalias{Prole2021} survey. The quantities \Mstar, \redshift\ and \recPhys\ are not observed but are constrained by the model.}
	\label{figure:model-summary}
\end{figure}

The posterior distribution implies \resultBeta\ and \resultAlpha, i.e. the prior distribution in $\alpha$ is recovered. Samples from the best-fitting model are shown in figure \ref{figure:model-summary}. The marginalised distributions in the model observed quantities \modelObsPars\ are consistent with observations, with KS statistic $p$-values$>$0.1. The uncertainties from the model fit are verified by resampling the best-fitting model with the same sample size as the observed dataset several times.

\indent The majority (\resultDwarfPercentage) of samples in the best-fitting model are dwarf galaxies, with \Mstar$\leq$$10^{9}$\Msun. The result indicates that the observations do not contain many galaxies with stellar masses lower than $\sim$$10^{7}$\Msun. Furthermore, \resultUDGPercentage\ of samples are UDGs according to the \cite{vanderBurg2016} definition. As expected, the majority of galaxies are local, with $z$$<$0.1.

\begin{figure}
	\includegraphics[width=1\linewidth]{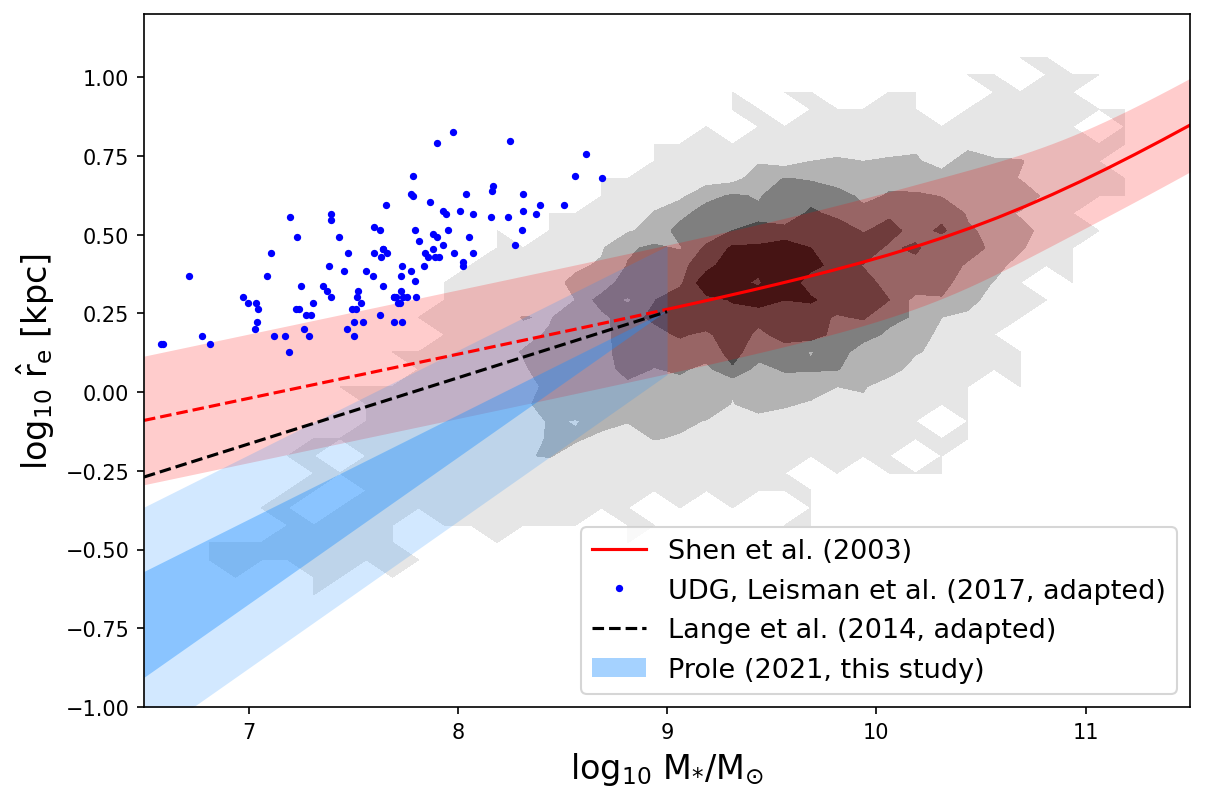}
	\centering
	\vspace*{-5mm}
	\caption{Stellar mass to circularised effective radius relation for late-type GAMA galaxies (greyscale), overlayed with the \protect\cite{Shen2003} model (extrapolated below $10^{9}$ \Msun) and \protect\cite{Lange2015} model, which has been converted into circularised effective radii assuming a mean axis-ratio of 0.7. The 1$\sigma$ confidence interval in the best-fit model is shown in dark blue, with the 1$\sigma$ scatter in light blue. H\RNum{1}-bearing UDGs from \protect\cite{Leisman2017} are also shown. The GAMA data are not used to fit the model.}
	\label{figure:mstar-rec}
\end{figure}

\indent As mentioned in $\S$\ref{section:modelColours}, the colour model adopted for dwarf galaxies introduces a systematic error in the model. The observed samples of \citetalias{Prole2021} and \cite{Leisman2017} suggest the mean of this distribution is at $(g-r)$=0.35. However, bluer galaxies tend to be of higher surface brightness at fixed stellar mass, meaning that assumptions based solely off LSB samples can be prone to bias. Making the mean bluer actually results in a steepening of the slope $\beta$, making the result more extreme, but this is disfavoured by the sizes of dwarf galaxies measured in the GAMA survey (figure \ref{figure:mstar-rec}). Adopting a redder mean value of $(g-r)$$\sim$0.4 \citep[as suggested by some models of UDG formation][]{Rong2017}, makes the \Mstar-\recPhys\ relation shallower, but this reduces the quality of the fit as quantified by the KS statistic. The net effect of varying the colour model is to introduce an additional uncertainty in $\beta$ of $\sim$0.05, which is incorporated into the final result: \resultBetaFinal.

\section{Discussion}
\label{section:discussion}

\subsection{Stellar mass - size relation}

The true slope of the mass - size relation has not been established in the dwarf galaxy regime because of surface brightness incompleteness, so a natural expectation would be for the slope measured in this work to be shallower than that of canonical models measured using higher surface brightness samples. However, it has not been established if extrapolations of the \cite{Shen2003} or \cite{Lange2015} models are appropriate for dwarf galaxies. In point of fact, it is clear from figure \ref{figure:mstar-rec} that such extrapolations already significantly over-estimate the sizes of dwarf galaxies present in the GAMA catalogue.

Indeed, the best-fitting model implies $\beta$ is significantly (3$\sigma$) steeper than would be expected from such extrapolations. Furthermore, the best-fitting model presented here appears to overlap with the mean trend of higher surface brightness dwarf galaxies in the GAMA catalogue (figure \ref{figure:mstar-rec}), meaning that the LSB galaxy population exists in the high-\recPhys\ tail of the mass-size relation and no significant alteration is required to the mean trend that would be measured directly from higher surface brightness dwarfs. This is consistent with the results of  \cite{Jones2018}, who showed H\RNum{1}-rich UDGs constitute a only a small fraction ($\leq$6\%) of detectable H\RNum{1}-bearing dwarf galaxies, with the majority being much smaller in size.

\indent This discrepancy between the best-fit model and extrapolated canonical models could be the reason why the semi-analytic model used by \cite{Jones2018}, who used the extrapolated \cite{Lange2015} relation, over-produces field UDGs. This idea is compounded by the position of the \cite{Leisman2017} UDGs in figure \ref{figure:mstar-rec}, which appear as 2$\sigma$ outliers from the mean relation. Incidentally, this is fully consistent with the alternative definition of UDGs introduced by \cite{Lim2020}, who defined them as significant outliers from empirical galaxy scaling relations.

\subsection{Ultra-diffuse galaxy size distribution}

\begin{figure}
	\includegraphics[width=0.85\linewidth]{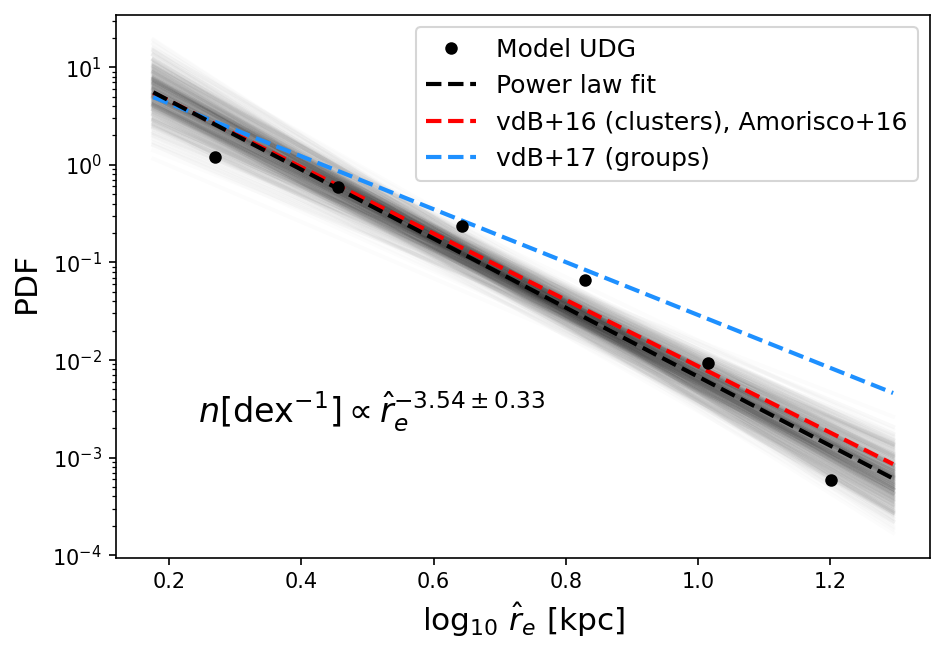}
	\centering
	\vspace*{-3mm}
	\caption{Power-law fit to model UDG size distribution in comparison with literature values \citep{vanderBurg2016, vanderBurg2017}. The new result is consistent with UDGs in clusters and those theoretically expected in the field \protect\citep{Amorisco2016}.}
	\label{figure:udg}
\end{figure}

The size distribution of UDGs can be obtained from the best fitting model after discounting selection effects (i.e. ignoring the recovery fraction $\epsilon$) and selecting only UDGs. A power-law model is fit to the resulting distribution of  \recPhys\ (figure \ref{figure:udg}). The best-fitting power law is $n [\mathrm{dex^{-1}}]\propto\recPhys^{\resultUdgPower}$, entirely consistent with the value measured for UDGs in clusters \citep{vanderBurg2016}, but inconsistent with the value measured for those in galaxy groups \citep{vanderBurg2017}.

The result is also consistent with the value predicted for field UDGs by \cite{Amorisco2016}, indicating higher than average halo spin parameters could be the dominant channel for UDG production. It is clear from figure \ref{figure:udg} that the power-law description may not provide the most precise representation of the UDG size distribution predicted by the model, which is qualitatively consistent with the small departure from the power-law visible in the results of \cite{Amorisco2016}.

\section{Data Availability}

\textit{The observed sample underlying this article is publicly available \citepalias{Prole2021}. The full code used to produce the results presented here is publicly available at \githuburl.}

\section{Acknowledgements}


\noindent I would like to thank the reviewer for their significant contribution to the quality of this article. D.P. acknowledges funding from an Australian Research Council Discovery Program grant DP190102448.


\bibliographystyle{mnras}
\bibliography{library.bib}

\end{document}